\documentstyle[preprint,revtex,eqsecnum]{aps}
\begin{document}
\tolerance 10000
\draft

\vskip 6 truecm
June, 1993 \hskip 9 truecm  Report LPQTH-93/07

\begin{center}
{\bf ON THE INTERNAL STRUCTURE OF THE SINGLET} \\
{\bf d$_{X^2-Y^2}$ HOLE PAIR IN AN ANTIFERROMAGNET}
\end{center}
\vspace*{0.4cm}
\author{Didier POILBLANC\cite{byline1}}
\vspace*{0.3cm}
\begin{instit}
\begin{center}
Groupe de Physique Th\'eorique,\\
Laboratoire de Physique Quantique,\\
Universit\'{e} Paul Sabatier,\\
31062 Toulouse, France
\end{center}
\end{instit}
\vspace*{0.2cm}

\receipt{\hskip 4truecm}

\begin{abstract}

Exact diagonalizations of two dimensional small t--J clusters
reveal dominant hole-hole correlations at distance $\sqrt{2}$ (ie
between holes on next nearest neighbor sites).
A new form of singlet $d_{x^2-y^2}$ pair operator is proposed to
account for the spatial extention of the two hole-bound pair beyond nearest
neighbor sites.

\end{abstract}
\pacs{PACS numbers: 75.40.Mg, 74.20.-z, 75.10.Jm, 74.65.+n}

Some time ago it has been proposed that the exchange of spin fluctuations
could lead to superconductivity in the singlet $d_{x^2-y^2}$
channel\cite{scalapino}.
Such mechanisms were recently discussed in connection with the
new copper oxyde superconductors. Occurence of d-wave pairing in strongly
correlated models
have been further supported by quantum Monte Carlo calculations\cite{qmc},
weak coupling approches\cite{pines} and recent finite size
scaling analysis\cite{26sites}.
The issue of pairing between holes moving in an antiferromagnetic
fluctuating background is of central importance
in the field of high-$T_c$ superconductivity.
Theoretical predictions\cite{nmrtheory} for NMR in a d-wave superconductor
seem to be consistent with the measured data\cite{nmrexp}.
Also consistent with d-wave pairing let us briefly mention recent
microwave experiments\cite{microwave} showing a linear temperature behavior of
the penetration depth and recent photoemission data\cite{photoemission}
revealing nodes in the gap function.

In standard notations the t--J Hamiltonian on a two dimensional (2D) reads
\begin{eqnarray}
{\cal H} = J \sum_{{\bf i},{\vec \epsilon}}
({\vec S}_{\bf i}\cdot {\vec S}_{\bf i+\vec \epsilon} - \frac{1}{4}
n_{\bf i} n_{\bf i+\vec \epsilon} )
-t \sum_{{\bf i},{\vec \epsilon},{\sigma}}
(c^{\dagger}_{{\bf i},{\sigma}}
 c_{{{\bf i}+{\vec\epsilon}},{\sigma}}
+ h.c. ) ,
\label {tj}
\end{eqnarray}
\noindent
where ${\vec S}_{\bf i}$ and $c_{{\bf i},\sigma}$ stand for the localized spins
and the itinerant holes respectively.

Numerical calculations of the binding energy between two holes on
$4\times 4$ \cite{manou,prelov1,jose1,hase},
$\sqrt{18}\times\sqrt{18}$ \cite{jose2},
$\sqrt{20}\times\sqrt{20}$ \cite{20sites} and
$\sqrt{26}\times\sqrt{26}$ \cite{26sites} square clusters suggested that
binding occurs for realistic values of the antiferromagnetic
exchange interaction ($0.3<J/t<0.6$)\cite{note}.
The bound state is found to be
a spin singlet with $d_{x^2-y^2}$ internal symmetry. As revealed by
a quasiparticle pole in the dynamical pair correlation
function\cite{pairsuscep,26sites} the bound pair
exhibits a coherent motion characteristic
of a unique (bosonic) particle of charge $+2e$ which leads to flux quantization
with a periodicity in flux of $\Phi_0/2$\cite{flux}.

The hole-hole correlations $C(R)$ in the two hole-ground state (GS) of the
$\sqrt{26}\times\sqrt{26}$ cluster are shown in Fig. 1
in good agreement with previous data on the $4\times 4$
cluster\cite{prelov1,hase} and
with approximate results on the same cluster obtained from a restricted
Hilbert space analysis\cite{xenophon}. Over a broad region, $0.2<J/t<0.9$,
the correlation at distance $\sqrt 2$ dominates. Strong correlations
are also found at distance 1 (i.e. for holes sitting on nearest neighbor sites)
which become in turn dominant for large J/t (unphysical). If the two holes
do form a bound state, as it is suggested by a finite size scaling
analysis\cite{26sites}, we expect the correlation $C(R)$ at small distance
R as well as the mean-hole separation\cite{26sites} to remain finite in the
thermodynamic limit.

Our purpose here is to define a proper pair operator which properly accounts
for the internal structure of the pair, with strong
density-density correlations at distance 1 {\it and} $\sqrt 2$.
A generalized form of a pair operator can be
defined as
\begin{eqnarray}
\Delta = \sum_R \frac{\big< \Psi_0^{N-2}\mid \Delta_R\mid \Psi_0^N\big>}{\big<
\Psi_0^N\mid  \Delta_R^\dagger \Delta_R\mid \Psi_0^N\big>}
\, \Delta_R ,
\label {pairoperator}
\end{eqnarray}
\noindent
where $\Psi_0^N$ and $\Psi_0^{N-2}$ are the N\'eel and two hole GS respectively
on the N site cluster and $\Delta_R$ creates a pair of holes on two sites
at a distance R.
The sum can be restricted to a small set of lattice distances
determined by the actual size of the pair.
Fig. 1 suggests that, in practice, it is sufficient to include
$R=1$ and $R=\sqrt 2$. The prefactors in (\ref{pairoperator})\cite{note2}
are chosen in
order to maximize the quasiparticle (QP) weight of the
pair \cite{dagotto,26sites},
\begin{eqnarray}
Z_{2h}= \frac{\mid\big< \Psi_0^{N-2}\mid \Delta\mid \Psi_0^N\big>\mid^2}{\big<
\Psi_0^N\mid  \Delta^\dagger \Delta\mid \Psi_0^N\big>}
=\sum_R \frac{\mid\big< \Psi_0^{N-2}\mid \Delta_R\mid
\Psi_0^N\big>\mid^2}{\big<
\Psi_0^N\mid  \Delta_R^\dagger \Delta_R\mid \Psi_0^N\big>} .
\label {weight}
\end{eqnarray}
\noindent
{}From Eq. (\ref{weight}) it is clear that the QP weight
increases with the number of pair operator $\Delta_R$ included
in the sum. Hence adding a contribution at $R=\sqrt 2$ could
significantly increase the weights $Z_{2h}$ calculated in Ref. \cite{26sites}
with the conventional $d_{x^2-y^2}$ BCS spin singlet operator for
nearest neighbor sites
$\Delta_1 = \sum_{\bf i} c_{{\bf i},\uparrow}
(  c_{{\bf i+{\hat x}},\downarrow} +
   c_{{\bf i-{\hat x}},\downarrow} -
   c_{{\bf i+{\hat y}},\downarrow} -
   c_{{\bf i-{\hat y}},\downarrow} )$.

Unlike for R=1, the pair operator at separation $\bf {\hat x + \hat y}$ can not
be written in a simple BCS form because of the $d_{x^2-y^2}$ character
of the GS wavefunction. Indeed, simple symmetry arguments
imply that the pair operator must be odd under reflections with respect to
the $x=\pm y$ directions.
Let us define $\Delta_{\sqrt 2}$ as
\begin{eqnarray}
\Delta_{\sqrt 2}=\sum_{\bf i}\{
(\vec S_{\bf {i+\hat y}}-\vec S_{\bf {i+\hat x}})\cdot
\vec T_{\bf i, \bf {i+\hat x+\hat y}}  -
(\vec S_{\bf {i-\hat x}}-\vec S_{\bf {i+\hat y}})\cdot
\vec T_{\bf i, \bf {i-\hat x+\hat y}}\}  ,
\label {nnnpairing}
\end{eqnarray}
\noindent
where $\vec T_{\bf i, \bf j}=\frac{1}{i}c_{\bf i,\sigma}
(\sigma_y\vec \sigma)_{\sigma\sigma^\prime}c_{\bf j,\sigma^\prime}$ is the
regular spin triplet pair operator\cite{note3}.
This expression is very similar to the pair function proposed by
Balatsky and Bon\c ca\cite{oddpairing} for the 1D t--J model in the context
of odd-gap pairing. However, we stress that (\ref{nnnpairing}) is, as far as
symmetry properties are concerned, entirely similar to $\Delta_1$ (spin
singlet, $d_{x^2-y^2}$ spatial symmetry) as schematically depicted in Fig. 2
whereas the odd-gap singlet operator of \cite{oddpairing} is {\it odd} under
space inversion, $\bf r\rightarrow -\bf r$ (p-wave character).

In summary, the hole-hole density correlation function
calculated exactly on a $\sqrt{26}\times\sqrt{26}$ cluster shows further
evidence in favor of a two hole $d_{x^2-y^2}$ bound state in the
fluctuating antiferromagnetic background. In order to include the
strong tendancy of the holes to stay together across the diagonal of a
plaquette a new form of pair operator is introduced which
couples the hole pair to the neighboring spins. Although its form
is very similar to pair field operators introduced in the context of
odd-gap pairing its symmetry properties are identical to the
symmetry properties of the conventional BCS pair operator at distance 1.

Support from the Centre de Calcul Vectoriel
pour la Recherche (CCVR), Palaiseau, France is acknowledged. Laboratoire
de Physique Quantique is Unit\'e Associ\'ee No. URA505 du CNRS.

\newpage
\bigskip
\centerline{FIGURE CAPTIONS}
\medskip

\noindent
{\bf Figure 1}

\noindent
Hole density correlations for 2 holes on a $\sqrt{26}\times \sqrt{26}$
cluster. The symbols corresponding to the various hole-hole separations are
shown on the plot.

\noindent
{\bf Figure 2}

\noindent
Schematic picture of the d-wave pair operator $\Delta_{\sqrt 2}$ around a
lattice site $\bf i$.
The arrows correspond to the triplet pair operators (oriented) and the
signs $\pm$ in parenthesis correspond to the phase factors of the spin
operators in (\ref{nnnpairing}).


\begin{thebibliography}{10}
\bibitem{note2} For a standart BCS pairing operator the normalization factor
$\big<\Psi_0^N\mid  \Delta_R^\dagger \Delta_R\mid \Psi_0^N\big>$ is simply
equal to the spin-spin correlation
$\big< \Psi_0^N\mid (1/4-{\vec S_{\bf i}}\cdot{\vec S_{\bf i+\bf r}})
\mid \Psi_0^N\big>$
(where $\bf r$ is such that $R=\mid\bf r\mid$).
\end{thebibliography}
\end{document}